\documentclass[10pt]{article}

\usepackage{amsmath}
\usepackage{amssymb}

\usepackage{graphicx}

\usepackage{cite}

\usepackage{color} 

\usepackage[labelfont=bf,labelsep=period,justification=raggedright]{caption}

\makeatletter
\renewcommand{\@biblabel}[1]{\quad#1.}
\makeatother

\date{}

\begin{document}

% Title must be 150 characters or less
\begin{flushleft}
{\Large
\textbf{Dynamical patterns of p53 states driven by $PML$}
}
\bigskip
% Insert Author names, affiliations and corresponding author email.
\\
Md. Jahoor Alam$^{1,2*}$, Eyad M. AlShammari$^1$ and R.K. Brojen Singh$^2$
\\
\bigskip
$^1$College of Applied Medical Sciences, University of Hail, Hail, Saudi Arabia.
\\
\bigskip
$^2$School of Computational and Integrative Sciences, Jawaharlal Nehru University, New Delhi-110067, India.
\\
\bigskip
$^{*}$Corresponding author, Dr. Md. Jahoor Alam, j.alam@uoh.edu.sa
\\
\end{flushleft}

\section*{Abstract}

Recent studies suggest that PML (Promyelocytic Leukemia) suppresses p53 protein degradation by inhibiting MDM2 protein in the nucleus\cite{bern}, and regulates a number of biological functions \cite{lan}. We modeled a PML-MDM2-p53 regulatory network by integrating p53-MDM2 autoregulatory model\cite{pro} and related PML pathways. We found that p53 dynamics switched at various dynamical states induced by PML which corresponds to different cellular states. Our results show clear transitions among the stabilized, damped and sustain states under different stress conditions induced by PML in the system. These states in p53 dynamics is the signature of existence of various cellular states and the phase transition like behaviour of these states involve various biological significance.

\bigskip
{\bf Keywords:} PML (Promyelocytic Leukemia), PML-MDM2-p53 network, Phase transition, Deterministic model.

\section*{Introduction}

PML is a well known tumor suppressor nuclear protein \cite{bern}. It has a small nuclear structure and present in almost all mammalian cells. It regulates many processes such as cell division, terminal differentiation of myeloid precursosr cells and neural progenitor cell differentiation \cite{ish}. It also acts as a transcription factor for many important regulatory proteins which helps in initiation of DNA damage response, DNA repair mechanism and organization of chromatin. A number of regulatory proteins are associated with PML such as SUMO1 (Small ubiquitin-related modifier 1), SUMO2 (Small ubiquitin-related modifier 2), MDM2(Mouse double minute 2), Tp53(tumor protein p53), CREBBP (CREB binding protein), RARA (retinoic acid receptor, alpha), ZFYVE9(zinc finger, FYVE domain containing 9), UBC(ubiquitin C), DAXX(death-domain associated protein), SP100 (SP100 nuclear antigen) etc\cite{ish,dou,wel,li}. PML is a key protein for generation of PML-nuclear bodies (which is also known as kremer bodies), nuclear domain 10 or PML oncogenic domains\cite{bor}. PML is very sensitive as well as responsive to various stress like DNA damage, viral infection, transformation, and oxidative stress etc \cite{reg}. Recent studies suggest that PML can suppress p53 tumour suppressor protein degradation by controlling MDM2 protein in the nucleus\cite{bern}.

p53 is a widely studied and an important protein in study of normal as well cancerous cells. It controls many important regulatory pathways such as cell cycle, apoptosis, DNA repair, cell differentiation etc\cite{lan,vou,lev}. It is responsive to the various stress like osmotic stress, oxidative stress, DNA damage by UV and IR, chromosomal abbration, chemical exposure, hormonal variability etc. In recent research reports various models have been proposed to shown the p53 activation through different stresses \cite{hor}. MDM2 is a oncogenic protein which acts as negative regulator of p53 protein. MDM2 protein acts as E3 ubiquitin ligase and voluntarily interact with p53 and forms p53-MDM2 complex and this leads to decrease in p53 concentration level within the cell\cite{bai,fin}. Several studies have been performed to show the p53 and MDM2 interaction and control mechanism of p53-MDM2 network. Sevaral attempt are done to understand the mechanism to control p53 inside the cell but still there is scope to understand the p53 restoration and the molecular event associted within the system. 

In the present work we designed and studied a p53-MDM2-PML integrated network by incorporating PML to the p53-Mdm2 regulatory network. In this model PML protein interacts with both p53 as well MDM2 protein and activates p53 and induces p53 expression within the cell which leads to increase in p53 concentration level within the cell. Several experimental and theoretical models have been proposed to study the PML interaction with p53-MDM2 network but still the temporal bahaviour and dynamical states of p53 induced by PML is not fully understood. The role of noise in the system is not systematically incorporated and investigated. We study some of these problems in this work in order to understand how does PML interferes cellular activities via p53. Our work is organized as follows, PML induced p53-MDM2 biochemical regulatory model and its molecular interaction is described in section 2, numerical results are presented with discussions in section 3 and some conclusions are drawn based on the results we obtained in section 4.           

\section*{Materials and Methods}

\subsection*{PML induced $p53-Mdm2$ regulatory model}

p53-MDM2 autoregulatory process is considered to be a very important regulatory process since it interferes most of the cellular activities both in normal as well as cancer cells \cite{lan,fin}. p53 is reported to be synthesized in cell with a rate constant $k_5$\cite{pro}. Mdm2 is a negative regulator of p53, and its translation from Mdm2\_mRNA is considered to be occured with a rate $k_1$. p53 acts as transcription factor which interacts with $Mdm2$ gene due to which transcription of $Mdm2$ gene into $Mdm2\_mRNA$ occurs with a rate $k_2$. The half life of $Mdm2\_mRNA$ is low which leads to the degradation of the $Mdm2\_mRNA$ which is considered to be occured with a rate constant $k_3$. The ubiqutination of the Mdm2 protein is taken to occur with a rate constant $k_4$. Mdm2 interacts with p53 protein in nucleus and forms $p53\_Mdm2$ complex with a rate constant $k_7$. Further, Mdm2 which act as ubiqiutin ligase which leads to the degradation of p53 concentration in cell with a rate $k_6$\cite{fin}. The dissociation of the $p53\_Mdm2$ complex is considered to be occured with a rate $k_8$. PML is considered as a nucleoprotein which is constantly expressed within the nucleus and varies in system to system with rate $k_{PML}$. PML is reported to interact with Mdm2 to form $PML\_Mdm2$ complex with a rate constant $k_{10}$. Furhter the degradation of Mdm2 occurs due to its interaction with PML, and is assumed to occur with a rate constant $k_{11}$. Moreover, PML changes the structural configuration of Mdm2 so that it is not available to interact with p53 and due to which p53 concentration increases in the cell. PML also directly interacts with p53 and forms $PML\_p53$ complex with a rate constant $k_{13}$. The interaction of PML and p53 leads to the phosphorylation of p53 protein on serine 20 \cite{lam,zen,mee}. This phosphorylation of p53 saves it from ubiquitation induced by Mdm2. Further, the dissociation of $PML\_p53$ complex takes place with a rate constant $k_{14}$. Moreover, degradation of the PML protein due to its shorter half life is reported to be taken place with a rate consrtant $k_{12}$ \cite{reg}. In Table 1, we listed molecular species associated with p53-MDM2-PML network. In Table 2, we listed the reaction channels, their rate constant and values taken in the simulation. The biochemical reaction network of the model is shown in Fig. 1.

Let the configurational state vector of the system we consider corresponding to state variables (molecular species in Table 1) be given by ${\bf x(t)}=\{x_1,x_2,...,x_n\}^T$, where, $T$ is the transpose of the vector and $n=7$
Based on the biochemical network shown in figure 1, we have translated the chemical reactions (Table 2) into a set of coupled ordinary differential equation using mass action law given by,
\begin{eqnarray}
\label{deter}
\frac{dx_1}{dt}&=& k_5-k_7x_1x_2+k_8x_3-k_{13}x_1x_5+k_{14}x_7\\
\frac{dx_2}{dt}&=& k_1x_4-k_4x_2+k_6x_3-k_7x_1x_2+k_8x_3-k_{10}x_5x_2\\
\frac{dx_3}{dt}&=& -k_6x_3+k_7x_1x_2-k_8x_3\\
\frac{dx_4}{dt}&=& k_2x_1-k_3x_4\\
\frac{dx_5}{dt}&=& k_{PML}-k_{10}x_5x_2+k_{11}x_6-k_{12}x_5-k_{13}x_1x_5\nonumber \\
&&+k_{14}x_7-k_{16}x_5x_6\\
\frac{dx_6}{dt}&=& k_{10}x_5x_2-k_{11}x_6\\
\frac{dx_7}{dt}&=& k_{13}x_5x_1-k_{14}x_7
\end{eqnarray}
where, $\{k_i\}$  $i=1,2,\dots,M (M=14)$ represent the sets of rate constants of the reactions listed in Table 2 and names of the molecular species listed in Table 1. We have used runge kutta 4th order algorithm for numerically solving the set of differential equations of our system. We developed our own coding simulation software which is written in fortran language \cite{pre}.  

\section*{Results and Discussion}

We present our numerical results as well as discussion on the results. We also tried to relate our results with possible biological processes arised due to PML interaction with p53-Mdm2 regulatory network.

\subsection*{$p53-Mdm2$ bio-chemical network activation}

The temporal behaviour of p53 protein for different values of PML creation rate constant $k_{PML}$ (Fig. 2) shows different states driven by PML. The small values of $k_{PML}$ ($\langle 0.0008$) could not able to activate from its stabilized state which corresponds to nearly normal state (Fig. 2 lowermost panel). However, for slight larger values of $k_{PML}$ ($\langle 0.002$), p53 dynamics exhibits damped oscillation for certain range of time (activated state) which depends on $k_{PML}$ and then maintains stabilized (come back to normal condition). Further, increase in $k_{PML}$ elongates the time of activation, and p53 dynamics become sustain oscillation state ($\sim 0.006$) showing strongest activation induced in the system. Further increased in $k_{PML}$ forces the p53 dynamics to damped oscillation and then reached stabilized state at larger concentration level (signature of apoptosis). Further, it is observed as the concentration level of PML increases which corrosponds to the rate constant of synthesis of PML within the cell, the temporal behaviour of p53 is also changes. This study suggests that as PML concentration increases the stress level in the cell also increases which is suppose to be due to molecular interaction of PML with p53-Mdm2 network. Further, It is also to be noted that as value of $k_{PML}$ increases from this particular range, the amplitude is also increases. This is probably due to higher stress generated by PML, which induces the p53 activation. This indicates that at this particular concentration of PML, the system is succumb to the stress and system is moving towards the apoptosis(cell death). Finally, when the concentration of PML is reached to be very high i.e at $k_{PML}$=0.05, we have got steady state level of p53. This suggest that, the system is succumb to stress and there is no further activation of p53 is observed and and cell goes for apoptosis(apoptosis)\cite{kub}. 

Similarly, we have analysed the temporal behaviour of Mdm2 protein as shown in Fig. 3 for same values of $k_{PML}$ as in the case of p53. Similar effect is observed but correspodingly in reverse manner as shown in p53 temporal behaviour which is supposed to be due to the sequestering effect of PML upon Mdm2 which is shown in Fig. 3.

A two dimensional plot of p53-Mdm2 network and its associated molecule is also shown in Fig. 4. In the first left hand Panel in Fig 3, the quantitave measurement of p53 verses Mdm2 which shown through correspoding configuration space for various values of $k_{PML}$. These results further verify the temporal behaviour discussed above.
        
\subsection*{Stability analysis of $p53$ and $Mdm2$}

Stability analysis for the system is shown in Fig. 5 and 6. We have analysed the stability of the system by two parameter, namely amplitude and maxima in the dynamics (p53 and Mdm2). In Fig. 5 the maxima of p53 and Mdm2 are found to be remain stable up to  certain values of $k_{PML}$, then it increases with respect to $k_{PML}$ values up to 0.006  and after which it decreases i.e in between $k_{PML}$=0.006 to 0.008. Further increasing $k_{PML}$ values become constant. This observation suggests that as PML concentration increases within the systems, the stress upon the system increases which leads to the increase in synthesis of p53 in the system which we can easily understand from the increase in maxima in p53 and Mdm2 and corresponding amplitudes (Fig. 6). We can also observe the amplitude death regimes (for small and large $k_{PML}$ values), and in between stress or activated regime.

\subsection*{Steady State Solutions for the systems}

The steady state solutions of the deterministic equations given by (1)-(7) corresponding to the concentrations of the molecular species can be obtained by taking, $d[x_i]/dt=0$, where $i=1,2,...,N$ $(N=7)$. Imposing these conditions we get seven differential equations involving seven variables corresponding to concentrations of the seven molecular species. Using these equations the steady state solution for $[x_1]$,  $[x_1]_{deter}^*$ can be obtained and is given below,
\begin{eqnarray}
\label{steady1}
[x_1]_{deter}^*=\sqrt{\frac{k_3k_4k_5}{k_1k_2k_7}\left(1+\frac{k_8}{k_6}\right)}
\end{eqnarray}
This steady state solution $[x_1]_{deter}^*$ gives us the behaviour of the p53 protein concentration when it is in extreme conditions and it is affected by various mechanisms. The equation (\ref{steady1}) indicates that p53 level is in fact influenced by various rate constants of various molecular species, their complex formation, decay and creation etc. If we make the rate of formation of MDM2, $k_1$ a variable then from this equation it is found that $[x_1]_{deter}^*$ is inversely proportional to $\sqrt{k_1}$ indicating the inhibiting activity of MDM2 to the steady state level of p53. The factor $k_8/k_6$ can be seen as the rate of dissociation of p53\_MDM2 per p53 decay rate i.e. a factor which indicate average amount of availability of p53 and MDM2 in the system. As $k_8/k_6$ increases more p53 and MDM2 are available in the system and vice versa.

We then solve for steady state solution of MDM2 i.e. $[x_2]^*_{deter}$ as explained above and is given by,
\begin{eqnarray}
\label{steady2}
[x_2]_{deter}^*=\sqrt{\frac{k_1k_2k_5}{k_3k_4k_7}\left(1+\frac{k_8}{k_6}\right)}
\end{eqnarray}
Similarly, $[x_2]_{deter}^*$ gives us the steady state level of MDM2 in extreme conditions and as in the case of p53, its steady state level is affected by other rate constants. However this steady state level is directly proportional to the $\sqrt{k_5}$ which is the rate of p53 protein formation. This means that the MDM2 steady state is strongly activated by p53 protein interaction in the network which is in agreement with the experimental results reported in qualitative sense.

Similarly, following the same procedure the steady state solutions of the variables $[x_3]_{deter}^*$ and $[x_4]_{deter}^*$ are obtained by solving the steady state equations and are given by, 
\begin{eqnarray}
\label{st}
[x_3]_{deter}^*=\sqrt{\frac{k_2k_4k_5}{k_1k_3k_7}\left(1+\frac{k_8}{k_6}\right)}
\end{eqnarray}
\begin{eqnarray}
\label{st1}
[x_4]_{deter}^*=\frac{k_5}{k_6}
\end{eqnarray}
From equation (\ref{st}) that $[x_3]_{deter}^*\propto\frac{1}{\sqrt{k_1}}$ indicating suppressing effect of MDM2 on MDM2\_mRNA, however, $[x_3]_{deter}^*\propto\sqrt{k_5}$ which reveals that p53 activates MDM2\_mRNA. Further, from equation (\ref{st1}) we have $[x_4]_{deter}^*\propto k_5$ indicating p53\_MDM2 is activated by p53. Now we solve for $\frac{[x_5]^*}{[x_6]^*}$ by relating to the rate constants of p53 and MDM2 as in the following,
\begin{eqnarray}
\frac{[x_5]_{deter}^*}{[x_6]_{deter}^*}=\frac{k_{11}}{k_{10}}\sqrt{\frac{k_3k_4k_7}{k_1k_2k_5\left(1+\frac{k_8}{k_6}\right)}}
\end{eqnarray}
At constant level of $[x_6]_{deter}^*$, the inversely proportional of $[x_5]_{deter}^*$ to $\sqrt{k_1}$ gives rise the suppressing activity of MDM2 to PML steady state level which is in agreement with the experimental results and predictions. 

\subsection*{Stress driven by PML and Mdm2 interaction}

PML interacts with p53 and Mdm2 allowing to change the state of the system. The interaction of PML and p53 is one important interaction which can change the fate of the system via p53. The interaction rate of PML and Mdm2, $k_{10}$ indicates the change in Mdm2 due to change in PML concentration level. The increase in $k_{10}$ allows the increase in amplitude of p53 as well as Mdm2 indicating increase in activation in the system (Fig. 7). Further, it is also observe that  as $k_{10}$ increases the amplitude death regime in small $k_{PML}$ regime (normal state regime) become shifted towards smaller $k_{PML}$ direction. It means that increase in $k_{10}$ the system is activated quicker. Similarly, the increase in $k_{10}$ extends the amplitude death regime. Surprisingly, this increase in $k_{10}$ needs larger values of $k_{PML}$ to reach apoptosis. Same scenario is obtained for Mdm2 case also.

\section*{Conclusion}
PML protein is found to be a tumour suppresser protein and it activates many protein to acheive cellular apoptosis. Although it directly interacts with both p53 and Mdm2, its behaviour is different for both the proteins. In the first case it activates the expression and in second case it modified the protein properties so that it can not negatively feedback upon previous one as reported in various experimental results. Our results support the stabilization effect of PML upon p53 and Mdm2. Simulation results clearly explain the temporal behaviour of the system with respect to stress driving the system to different states, stabilized, damped and sustain oscillations. 

The interaction of PML and Mdm2 enhances the activation of the system and resists the system to go to apoptosis. Similarly, interaction of PML with other interaction partners may influence in the cellular stress. The increase in PML level in cellular system may trigger cancer in some cases which is needed to be investigated further.

%\section*{Author's contributions}

%\section*{Acknowledgments}

%\section*{References}

\newpage

\section*{Tables}

\begin{table*}
\begin{center}
{\bf Table 1 - List of molecular species} 
\begin{tabular}{|l|p{2.5cm}|p{5cm}|p{2cm}|}
 \hline \multicolumn{4}{}{} \\ \hline

\bf{ S.No.}    &   \bf{Species Name}    &    \bf{Description}           &  \bf{Notation}     \\ \hline
1.             &    $p53$               & Unbounded $p53$ protein       &  $x_1$ \\ \hline
2.             &    $Mdm2$              & Unbounded $Mdm2$ protein      &  $x_2$ \\ \hline
3.             &    $Mdm2\_p53$         & $Mdm2$ with $p53$ complex     &  $x_4$  \\ \hline\
4.             &    $Mdm2\_mRNA$        & $Mdm2$ messenger $mRNA$       &  $x_3$  \\ \hline
5.             &    $PML$            & Unbounded $PML$ protein     &  $x_5$  \\ \hline
6.             &    $Mdm2\_PML$            &  $Mdm2\_PML$ Complex       &  $x_6$  \\ \hline
7.             &    $p53\_PML$            &  $p53\_PML$ complex  &  $x_7$  \\ \hline

\end{tabular}
\end{center}
\end{table*}

%\newpage

\begin{table*}
\begin{center}
{\bf Table 2 List of chemical reaction, kinetic law and their rate constant} 
\begin{tabular}{|l|p{2.5cm}|p{3.5cm}|p{2.5cm}|p{3cm}|p{2cm}|}
\hline \multicolumn{6}{}{}\\ \hline

${\bf S.No}$ & ${\bf Reaction}$ & ${\bf Name of the process}$ & ${\bf Kinetic Law}$ & ${\bf Rate Constant}$ & ${\bf References}$ \\ \hline
1 & $x_4\stackrel{k_{1}}{\longrightarrow}x_4+x_2$ & Mdm2 synthesis & $k_1 \langle x_4\rangle$ & $4.95\times 10^{-4}{sec}^{-1}$ & \cite{pro,fin}.\\ \hline
2 & $x_1\stackrel{k_{2}}{\longrightarrow}x_1+x_4$ & $Mdm2\_mRNA$ synthesis & $k_2 \langle x_1\rangle$ & $1.0\times 10^{-4}{sec}^{-1}$ & \cite{pro,fin}.\\ \hline
3 & $x_4\stackrel{k_{3}}{\longrightarrow}\phi$ & $Mdm2\_mRNA$ decay & $k_3 \langle x_4\rangle$ & $1.0\times 10^{-4}{sec}^{-1}$ & \cite{pro,fin}.\\ \hline
4 & $x_2\stackrel{k_{4}}{\longrightarrow}\phi$ & Mdm2 degradation & $k_4 \langle x_2\rangle$ & $4.33\times 10^{-4} sec{-1}$ & \cite{pro,fin}. \\ \hline
5 & $\phi\stackrel{k_{5}}{\longrightarrow}x_1$ & p53 synthesis & $k_5$  & $0.078{sec}^{-1}$ & \cite{pro,fin}.\\ \hline
6 & $x_3\stackrel{k_{6}}{\longrightarrow}x_2$ & $Mdm2\_p53$ degradation & $k_6 \langle x_3\rangle$ & $8.25\times 10^{-4}{sec}^{-1}$ & \cite{mee,rod}. \\ \hline
7 & $x_1+x_2\stackrel{k_{7}}{\longrightarrow}x_3$ & $Mdm2\_p53$ synthesis & $k_7 \langle x_1\rangle\langle x_2\rangle$ & $11.55\times 10^{-4}{sec}^{-1}$ & \cite{pro}. \\ \hline
8 & $x_3\stackrel{k_{8}}{\longrightarrow}x_1+x_2$ & $Mdm2\_p53$ dissociation & $k_8 \langle x_3\rangle$ & $11.55\times 10^{-6}{sec}^{-1}$ & \cite{pro}. \\ \hline
9 & $\phi\stackrel{k_{PML}}{\longrightarrow}x_5$ & PML creation & $k_{PML}$ & $5.0\times 10^{-3}{sec}^{-1}$ & \cite{rod}. \\ \hline
10 & $x_2+x_5\stackrel{k_{10}}{\longrightarrow}x_6$ & $PML\_Mdm2$ complex formation & $k_{10} \langle x_2\rangle\langle x_5\rangle$ & $2.0\times 10^{-3}{sec}^{-1}$ & \cite{rod}. \\ \hline
11 & $x_6\stackrel{k_{11}}{\longrightarrow}x_5$ & Decreasing of Mdm2 & $k_{11} \langle x_1\rangle$ & $3.3\times10^{-5} sec^{-1}$ & \cite{rod}. \\ \hline
12 & $x_5\stackrel{k_{12}}{\longrightarrow}\phi$ & PML decay & $k_{12} \langle x_5\rangle$ & $5.0\times 10^{-1}{sec}^{-1}$ & \cite{rod}.\\ \hline
13 & $x_1+x_5\stackrel{k_{13}}{\longrightarrow}x_7$ & $p53\_PML$ complex formation & $k_{13} \langle x_1\rangle\langle x_5\rangle$ & $1.0\times 10^{-6}{sec}^{-1}$ & \cite{kni,luo}. \\ \hline
14 & $x_7\stackrel{k_{14}}{\longrightarrow}x_1+x_5$ & dissociation of $p53\_PML$ complex & $k_{14} \langle x_7\rangle$ & $1.0\times 10^{-6}{sec}^{-1}$ & \cite{kni,luo}. \\ \hline

\end{tabular}
\end{center}
\end{table*}

\newpage

\section*{Figure Legends}

\subsection*{Figure 1 - PML induced $p53-Mdm2$ biochemical regulatory model}
The schematic diagram of stress p53-Mdm2-PML model.
\subsection*{Figure 2a - Effect of PML on p53 protein}
Temporal behaviour of p53 due to the switching of PML protein at various concentration level i.e 0.04,0.01,0.009,0.006,0.003,0.001,0.0005,0.0008.
\subsection*{Figure 2b - Effect of PML on Mdm2 protein}
Temporal behaviour of Mdm2 due to the switching of PML protein at various concentration level i.e 0.04,0.01,0.009,0.006,0.003,0.001,0.0005,0.0008.
\subsection*{Figure 3 - Two dimensional recurrence plot }
Two dimensional recurrence plots between ($p53-Mdm2$), ($p53-PML$), ($Mdm2-PML$), ($p53\_Mdm2-PML$), ($Mdm2\_mRNA-PML$)for different values of rate constants $k\_PML$, i.e. 0.0005, 0.001, 0.003, 0.006, 0.009, 0.01 and 0.03 respectively.
\subsection*{Figure 4 - Stability Analysis }
Plots for the stability analysis.In the upper panel stability of p53(in magenta) and Mdm2(in blue) is shown in terms of variablity in amplitude with respect to time. In lower panel stability of p53(in magenta) and Mdm2(in blue) is shown in terms of variablity in time period with respect to time.

\begin{figure*}
\label{}
\begin{center}
\includegraphics[height=20cm,width=16.0cm]{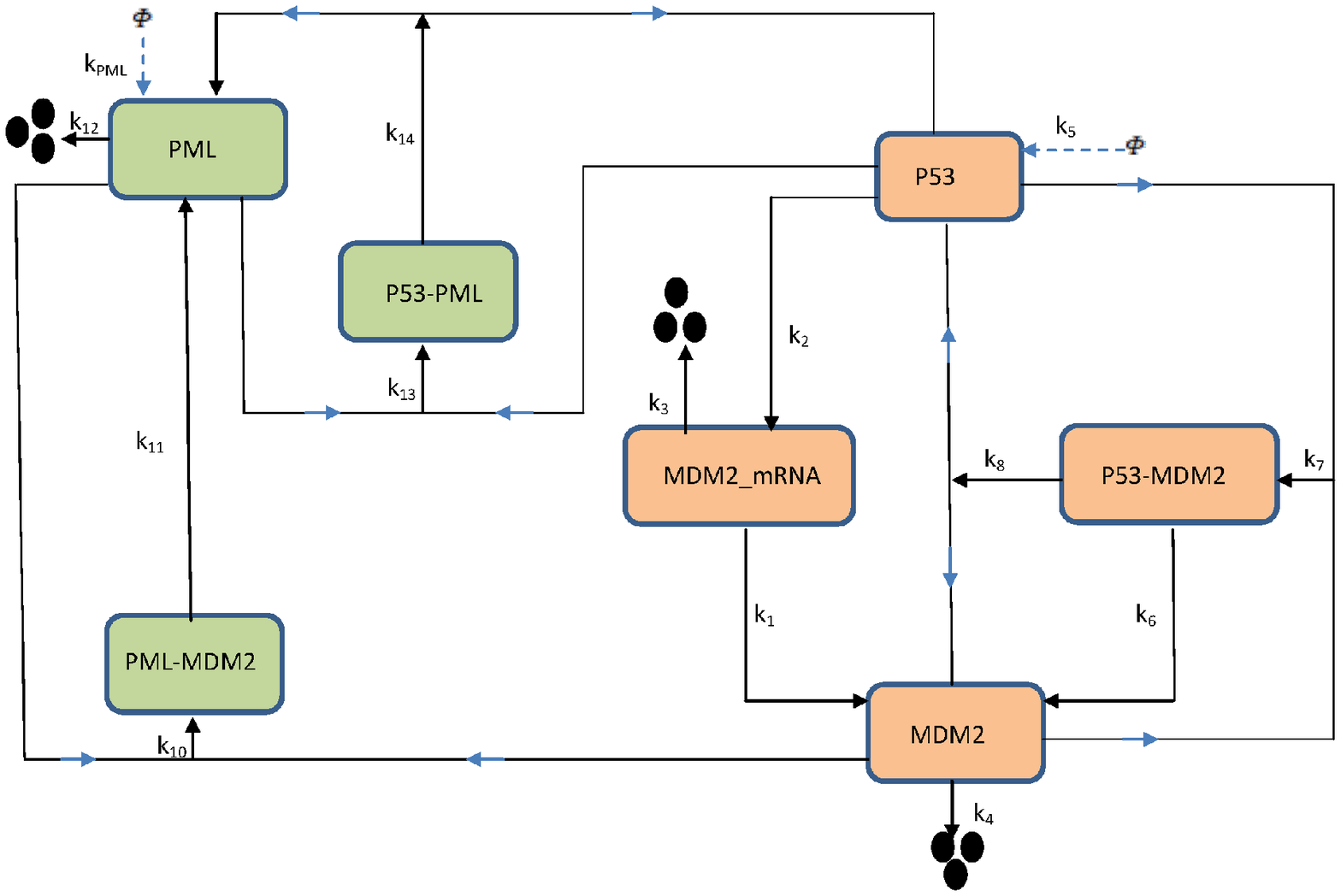}
\caption{} 
\end{center}
\end{figure*}

\begin{figure*}
\label{}
\begin{center}
\includegraphics[height=320 pt,width=11.0cm]{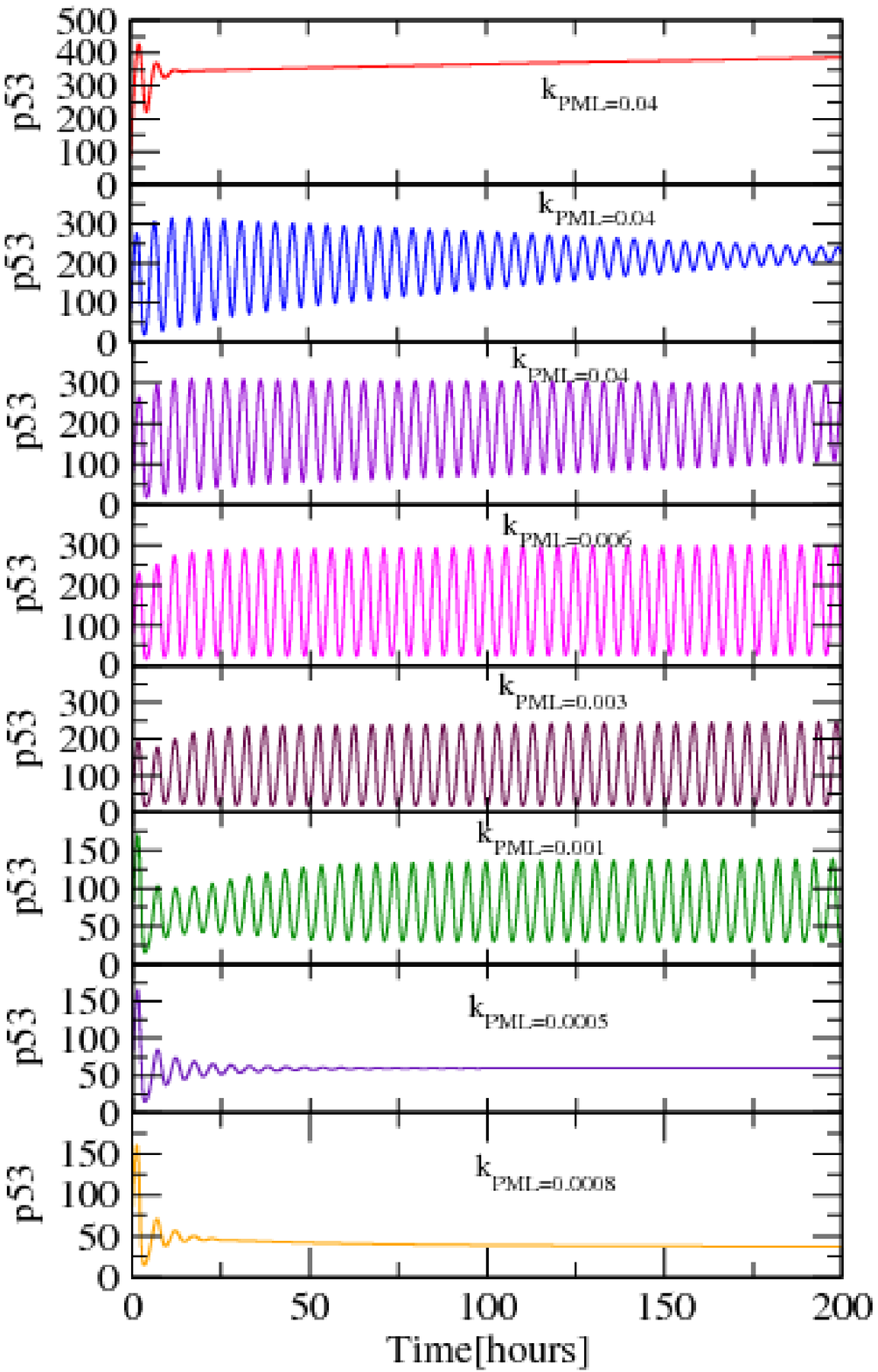}
\caption{} 
\end{center}
\end{figure*}

\begin{figure*}
\label{}
\begin{center}
\includegraphics[height=320 pt,width=11.0cm]{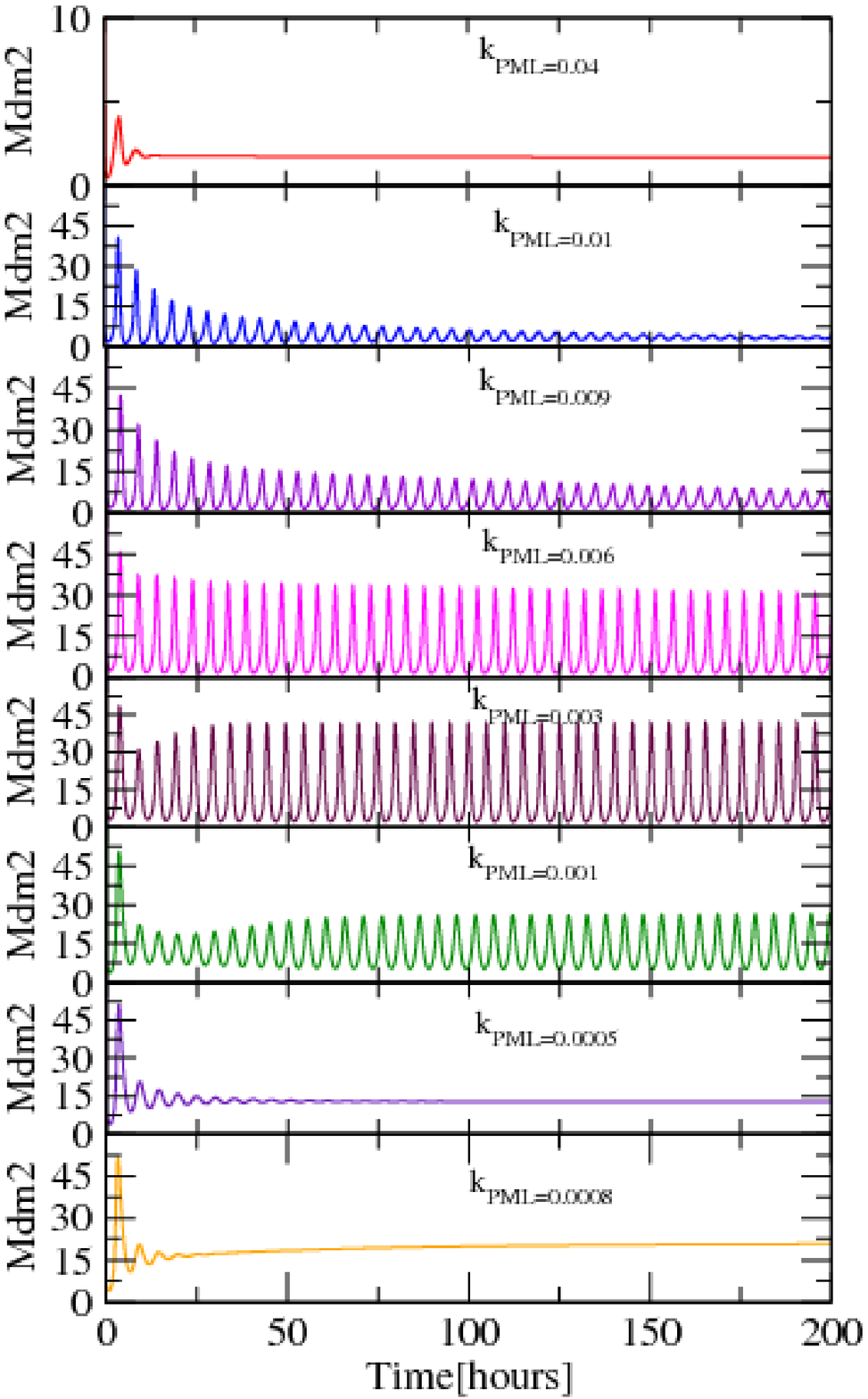}
\caption{} 
\end{center}
\end{figure*}

\begin{figure*}
\label{}
\begin{center}
\includegraphics[height=320 pt,width=11.0cm]{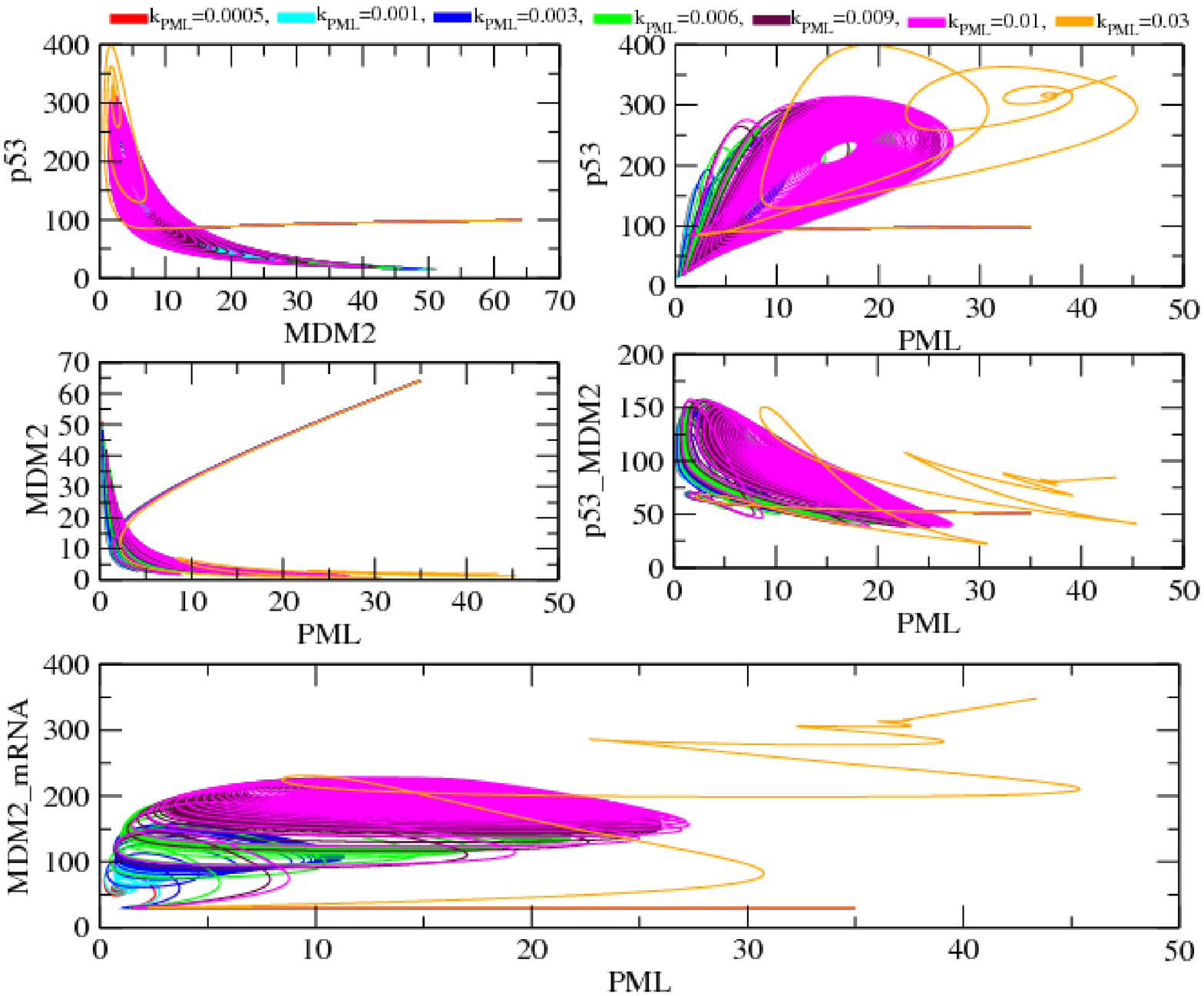}
\caption{} 
\end{center}
\end{figure*}

\begin{figure*}
\label{}
\begin{center}
\includegraphics[height=12cm,width=12.0cm]{figure4.eps}
\caption{} 
\end{center}
\end{figure*}

\begin{figure*}
\label{}
\begin{center}
\includegraphics[height=10cm,width=10.0cm]{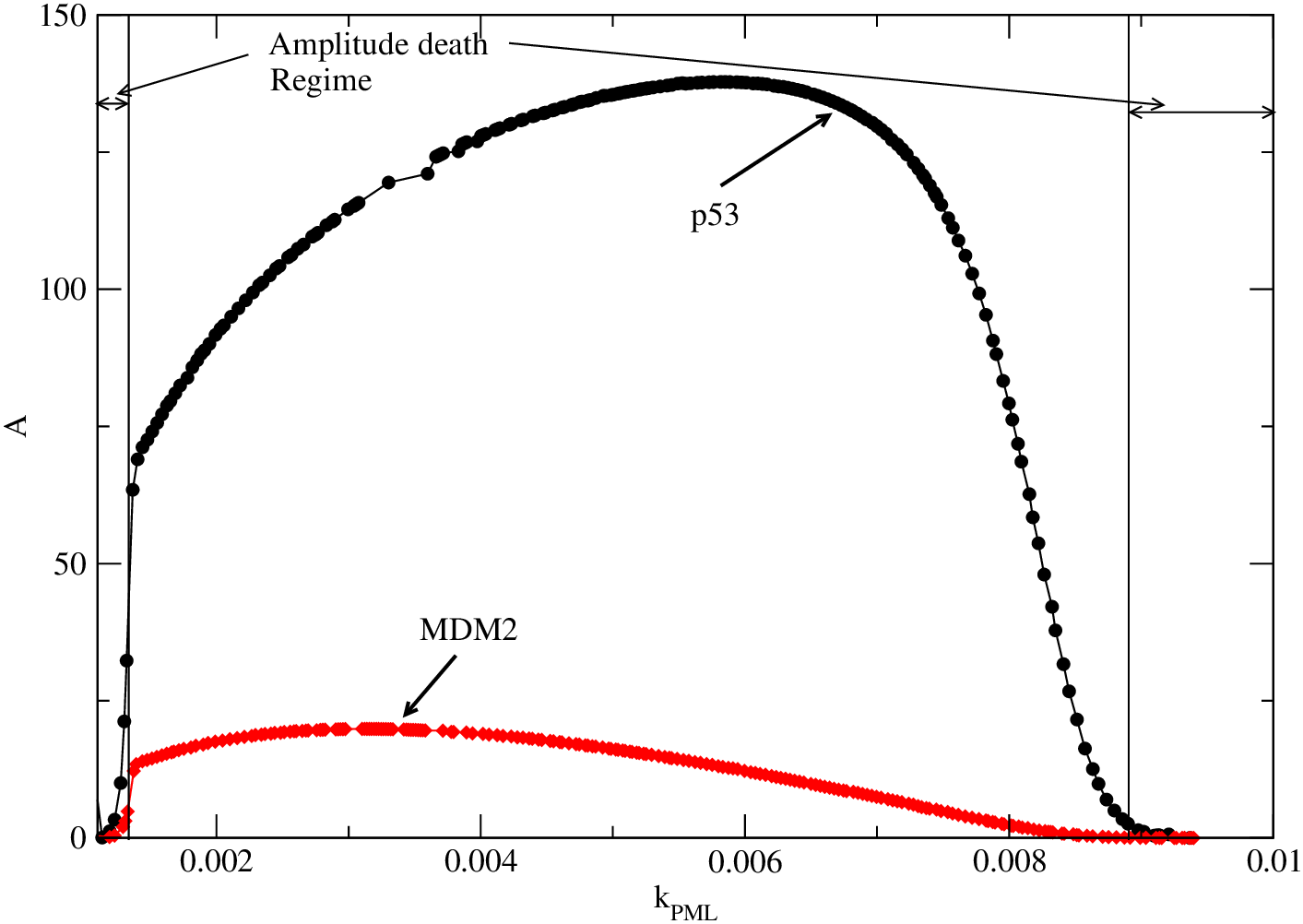}
\caption{} 
\end{center}
\end{figure*}

\begin{figure*}
\label{}
\begin{center}
\includegraphics[height=14cm,width=12cm]{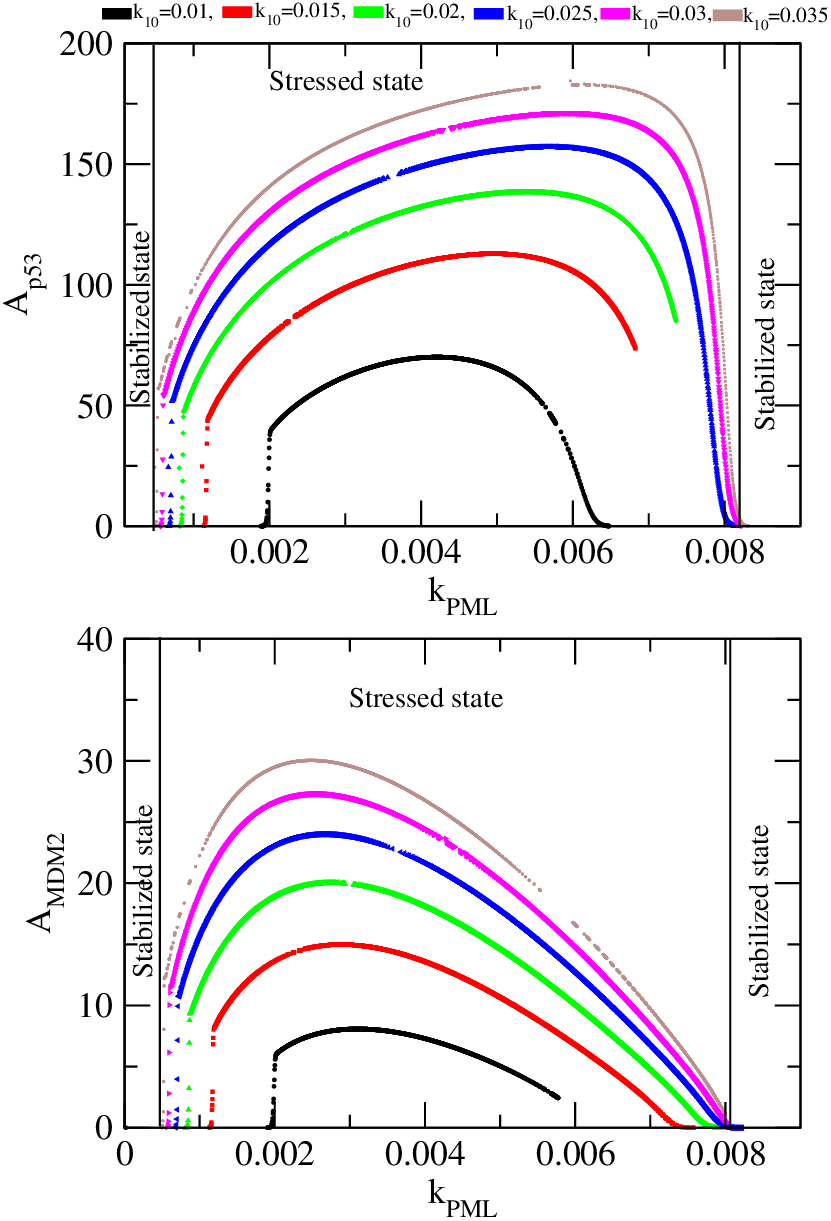}
\caption{} 
\end{center}
\end{figure*}

\end{document}